\documentclass[letterpaper,twocolumn,showpacs, %preprint, %
  amsmath,amssymb,showkeys, floatfix, prl]{revtex4}

\usepackage[dvips]{color,graphicx}
\usepackage[english]{babel}
\usepackage{ae}
\usepackage{aecompl}
\usepackage{dcolumn}% Align table columns on decimal point

\newcommand*{\prom}[1]{\ensuremath{\langle #1 \rangle}}
\newcommand*{\kb}{\ensuremath{k_{B}}}

\begin{document}

  % \preprint{APS/123-QED (PREPRINT)}

  \title{
    The Single Histogram Method and the Quantum Harmonic %
    Oscillator: Accuracy Limits
  }
  % \title{
  %   An Application of the Single Histogram Method to the Quantum
  %   Harmonic Oscillator
  % }

  \author{W.F. Oquendo}
  \email[email at: ]{woquendo@gmail.com}
  \author{J.D. Muñoz}%
  \email[email at: ]{jdmunozc@unal.edu.co}
  \affiliation{%
    Grupo Simulación de Sistemas Físicos, Departamento de Física\\
    Universidad Nacional de Colombia, Carrera 30 No. 45 - 03, Ed. 404,
    Of. 348, Bogotá, Colombia.
  }%

  \date{\today}

  \begin{abstract}
    In a recent work, M.~Troyer, F.~Alet and S.~Wessel~\cite{brazilean}
    proposed a way to extend histogram methods to quantum systems in
    the World Line Quantum Monte Carlo (WLQMC) formulation. The
    strategy, also proposed in \cite{josedaniel}, allows to compute
    quantum averages on a narrow temperature range from a single Monte
    Carlo run at fixed temperature. This is achieved by fixing $N$,
    the number of 
    temporal divisions in the Trotter-Suzuki expansion of  WLQMC, and by
    changing $\epsilon$$=$$1/(N \kb T)$. In this work we apply this strategy
    to construct a single histogram Monte Carlo method for a
    canonical ensemble of one-dimensional quantum harmonic oscillators
    and we explore its accuracy limits. We obtain that fixing $N$
    imposses a limit of minimal temperature to the properly performance
    of the method, which is $T_{min}$$=$$1.9(2)N^{-0.80(6)}$ in our
    example. This limit is a consequence of the fact that the
    Trotter-Suzuki expansion fails for large $\epsilon$ values, and,
    therefore, should be taken into account in all applications of this
    histogram method for quantum systems.
  \end{abstract}

  \pacs{02.70.Ss, 02.70.Tt, 02.70.Uu, 05.10.-a, 05.10.Ln}

  \keywords{Quantum Monte Carlo; Histogram Method; Computational Physics}

  \maketitle

  \section{\label{sec:intro}Introduction}
  Histogram Monte Carlo methods~\cite{josedaniel,binder,landau} are 
  powerful and robust
  techniques to investigate classical systems, both with discrete or
  continuous degrees of freedom. Such methods have many advantages over 
  traditional Monte Carlo ones. For example, one can obtain mean values
  of thermodynamical quantities over a broad range  of
  temperatures by using data from a single run.
  Also it is possible to compute thermodynamic potentials, like the free
  energy, that are not accessible by other methods.
  All these methods bases on defining a density
  of states, let say $g(E)$, i.e. the number or states 
  with the same energy. Once obtained, $g(E)$ allows to compute averages like
  the mean energy and the specific heat at any desired
  temperature. Therefore, the problem of averaging over all states is
  transformed into a problem of counting to find $g(E)$. 
  Different histogram methods differ in the way they obtain the density of
  states.  

  There are histograms methods for analysis and histograms methods for
  sampling. Some examples of the first ones are the Single Histogram Method
  (SHM), first proposed by Z.~W.~Salsburg in 1959~\cite{salsburg} and 
  popularized by A.~M.~Ferrenberg and
  R.~H.~Swendsen in 1988~\cite{ferrenberg}; the Broad Histogram Method
  (BHM) formulated in 1996 by P.~C.~M.~de~Oliveira, T.~J.~P.~Penna and
  H.~J.~Hermann~\cite{oliveira}, and the Transition Matrix Monte
  Carlo method (TMMC) proposed by J. S. Wang, T. K.
  Tay and R. H. Swendsen in 1998~\cite{tmmc0,tmmc1}, which reduces to
  BHM when a single energy
  jump is chosen. In the second class we find, for
  example, the Multicanonical Method, proposed by B.~A.~Berg\cite{berg} 
  (also formulated by J.~Lee as the Entropic Sampling method\cite{lee}), the
  Flat-Histogram method of Wang and Lee~\cite{flathisto1, flathisto2} and the
  Wang-Landau method~\cite{wanglandau}. These methods take samples
  uniformly distributed on the energy axis.

  Recently, M.~Troyer, F.~Alet and S.~Wessel proposed a way to extend histogram 
  Monte Carlo methods
  to canonical emsembles of quantum systems, in the world line quantum 
  Monte Carlo (WLQMC)
  formulation~\cite{brazilean, suizosletter}. In contrast with previous
  proponsals \cite{martin,tmmcq},  this idea, 
  also proposed in \cite{josedaniel} 
  resembles the full power of the clasical 
  histogram methods, since it is able, for instance, to calculate mean
  values on a narrow temperature range from a single Monte Carlo run at
  a fixed temperature (SHM). The idea bases on the observation that
  interactions in time and space in the equivalent classical system of WLQMC are
  weighted by the temperature in different ways, as
  we will illustrate below. Thus, it is useful to 
  define two new quantities, $k_1$ and $k_2$, which summarize spatial and
  temporal interactions, and also a density of states 
  $g(k_1, k_2)$ on them. Such density can be computed by
  using any histogram method and allows to obtain averages at different
  temperatures from a single run. This strategy has been used by
  Troyer, Alet and Wessel  to investigate the three-dimensional quantum
  Heisenberg model~\cite{brazilean}, with excellent results. 

  In this work we illustrate this strategy by studying a
  much more simple system: a canonical ensemble of non-coupled
  one-dimensional quantum harmonic oscillators. By doing so, interesting
  aspects of the method like its range of accuracy and its dependence
  on the simulation parameters will be
  explored. The article is organized as follows.
  First, the World Line Quantum Monte Carlo method is revised.
  Next, the quantities $k_1$, $k_2$ and $g(k_1, k_2)$ for a quantum
  single-particle system are defined, and it is shown how this density of
  states allows to compute expectation values for the kinetic, potential
  and total energy operators at any desired temperature.
  Then, we implement the SHM for the one-dimensional quantum harmonic
  oscillator, and we investigate how the temperature range of accuracy 
  is affected by the number of divisions on the axis of
  imaginary time.
  Finally, we summarize our main results and conclusions. 

  \section{\label{sec:wlqmc}The World Line Quantum Monte Carlo method}
  The World Line Quantum Monte Carlo method (WLQMC) 
  was formulated in 1982 by
  J.~E.~Hirsch, R.~L.~Sugar, D.~J.~Scalapino
  and R.~Blankenbecler~\cite{wlqmc}.
  It has been used to study one particle systems as well as many particle
  systems, like interacting bosons and system of strong correlated
  fermions. The WLQMC employs path integrals
  to change the partition function of a  quantum system to the partition
  function of an equivalent classical system in one higher
  dimension~\cite{lineas}. To illustrate the method, let us consider a
  quantum system with a separable Hamiltonian
  \begin{equation}\label{equ:genehami}
    \Hat H  = \Hat T(\Hat P) + \Hat V(\Hat X) = 
    \frac{\Hat P^2}{2m} + \Hat V(\Hat X),
  \end{equation}
  where $\Hat V(\Hat X)$ and $\Hat T = \Hat P^2/2m$ are the potential and 
  kinetic energy operators respectively. In the canonical ensemble, the
  partition function
  is~\cite{reichl}
  \begin{equation}\label{equ:genepartfunc}
    Z = \mathsf{Tr} [e^{-\beta \Hat H}],
  \end{equation}
  where $\beta = \frac{1}{\kb T}$ is the imaginary time, $\Hat H$ is
  given by \eqref{equ:genehami} and $\kb$ is the Boltzmann constant.

  Since $\Hat X$ and $\Hat P$ do not commute, the
  exponential in \eqref{equ:genepartfunc} cannot be separated into a 
  product of exponentials. Therefore, the imaginary time $\beta$ is written
  as $\beta = N\epsilon$, and a Trotter--Suzuki approximation is 
  used~\cite{suzuki1, suzuki2,suzuki3} %
  to write the partition function \eqref{equ:genepartfunc} as
  \begin{equation}\label{equ:partfunctr}
    Z = \lim_{\substack{N\to\infty\\ \epsilon\to 0}} Z_{tr} = 
    \lim_{\substack{N\to\infty\\ \epsilon\to 0}}
    \mathsf{Tr}[(e^{-\epsilon\Hat T}e^{-\epsilon\Hat V(\Hat X)})^N].
  \end{equation}
  To compute \eqref{equ:partfunctr}, the trace
  is expressed as a sum over a
  complete set of position eigenstates, and a new completeness relation is
  introduced between each of the $N$ factors in that expression,
  obtaining
  \begin{align}
    Z_{tr} &=\left( \frac{m}{2\pi\epsilon} \right)^{N/2}\int \mathcal{D}x\
    \exp[-\epsilon H(x)],\label{equ:zetatr}\\
    H(x) &= \sum_{l=0}^{N-1}V(x_l) + \frac{1}{2}m\sum_{l=0}^{N-1}\left(
    \frac{x_l - x_{l+1}}{\epsilon} \right)^2,\label{equ:hamiclas}
  \end{align}
  where $\mathcal{D}x= dx_0 dx_1 \dotsm d x_{N-1}$. The partition
  function for the quantum system is now expressed as a multidimensional
  integral over classical variables.  The expression~\eqref{equ:hamiclas} can
  be interpreted as the energy of a chain of $N$ classical oscillators with
  potential energy $V(x)$ coupled through springs of 
  constant $k=m/\epsilon^2$.
  In this case, $x = \{ x_0, x_1, \dotsc, x_{N-1}\}$ will denote 
  a specific configuration of this classical system. 

  By following the same procedure described above, one obtains
  \begin{equation}
    \prom{\Hat V}_{tr} = \frac{\int \mathcal{D}x\ V(x_0) \exp[-\epsilon
        H(x)]}{\int \mathcal{D}x\  \exp[-\epsilon H(x)]}, 
    \label{equ:prompote}
  \end{equation}
  \begin{equation}
    \prom{\Hat T}_{tr} = \frac{1}{2\epsilon} - \frac{m}{2\epsilon^2}
    \frac{\int \mathcal{D}x\ (x_0 - x_1)^2 \exp[-\epsilon H(x)]}
         {\int \mathcal{D}x\ \exp[-\epsilon H(x)]},
         \label{equ:promkine}
  \end{equation}
  for the expectation values of the potential and kinetic energy
  operators respectively,  where $x_0$ and $x_1$ can be replaced 
  by any pair $(x_i, x_{i+1})$ of neighboring classical oscillators. 
  For the total energy we have $\prom{\Hat H}_{tr} = \prom{\Hat T}_{tr} +
  \prom{\Hat V(\Hat X)}_{tr}$.

  Expressions~\eqref{equ:prompote} and \eqref{equ:promkine} 
  show that the expectation values for quantum
  operators
  are now expressed as multidimensional integrals over the configuration
  space of the classical system. This integral can be computed by using standard
  Monte Carlo methods. For example, we can use the Metropolis algorithm 
  to compute the total energy average $\prom{\Hat H}_{tr}$ 
  as follows. First, an oscillator $i$ in configuration $x$ is chosen at
  random and its 
  position $x_i$ is changed to a new value $x_i + r$, where $r$ is a uniformly
  distributed random number in $[-1,1]$. Next, this new configuration $x'$
  is accepted with rate $A(x' \mid x) = \min\{1,e^{-\epsilon (H(x') - H(x))} \}$.
  Finally, energy averages are computed as arithmetic means over the sample set.

  The temperature can be fixed by varying either $N$ or
  $\epsilon$, since $\beta = 1/\kb T = N\epsilon$. When $N$ is
  fixed and $\epsilon$ varies to fix the temperature, errors in the
  Trotter--Suzuki approximation are greater at low temperatures.

  \section{\label{sec:shmoacu} Histogram Methods for Quantum Systems in the WLQMC
    formulation} 
  For a classical system in the canonical ensemble, the average energy at
  temperature $T$ is given by
  \begin{equation}\label{equ:classenerprom}
    \prom{E}_T = \frac{\sum_x E_x e^{-\frac{E_x}{\kb T}}}{\sum_x 
      e^{-\frac{E_x}{\kb T}}} = \frac{\sum_E g(E)E
      e^{-\frac{E}{\kb T}}}{\sum_E g(E) e^{-\frac{E}{\kb T}}}, 
  \end{equation}
  where $g(E)$ is the number of configurations with the same energy $E$. 
  If $g(E)$ is known, one can use~\ref{equ:classenerprom} to obtain $\prom{E}_T$
  as a continuous function of temperature.  So, the problem of summing over all
  states has been transformed into a problem of counting to find $g(E)$.

  In the classical SHM, samples are taken with probability proportional to
  $e^{-E_x/\kb T}$ (canonical distribution) at fixed temperature $T$, and a
  histogram of visits $\mathcal{V}(E)$ is accumulated as the number of
  samples with energy $E$. From such histogram, the density of states $g(E)$
  is approximated as
  \begin{equation}\label{equ:classgeshm}
    g(E) \simeq \frac{\mathcal{V}(E)}{\mathcal{N}}\exp[E/\kb T],
  \end{equation}
  where $\mathcal{N}$ is the total number of samples. By
  replacing~\eqref{equ:classgeshm} in~\eqref{equ:classenerprom}, we can
  obtain averages at temperatures $T' \ne T$.

  To extend histogram methods to quantum systems, one has to define a
  density of states. As proposed in~\cite{brazilean,josedaniel}, this is
  achieved by defining two quantities in the WLQMC classical system,
  which resumes the interactions in space and time directions:
  \begin{equation}
    k_1 = \sum_{l=0}^{N-1} V(x_l), \ \ 
    k_2 = \sum_{l=0}^{N-1}(x_l - x_{l+1})^2\label{equ:kunodos},
  \end{equation}
  with periodic boundary conditions. The partition
  function~\eqref{equ:zetatr} and the 
  averages~\eqref{equ:prompote} and~\eqref{equ:promkine}
  can be written in terms of $k_1$ and $k_2$ as
  \begin{equation}
    Z_{tr} =\left( \frac{m}{2\pi\epsilon} \right)^{N/2}\int dk_1 dk_2\ 
    g(k_1, k_2) e^{-\epsilon H(k_1, k_2)},\label{equ:zetatrk1k2}
  \end{equation}
  \begin{equation}
    \prom{\Hat V}_{tr} = \frac{1}{N}\frac{  \int dk_1 dk_2\  g(k_1, k_2) k_1\ 
      e^{-\epsilon
        H(k_1, k_2)}}{\int   dk_1 dk_2\   g(k_1, k_2) e^{-\epsilon H(k_1, k_2)}}, 
    \label{equ:prompotek1k2}
  \end{equation}
  \begin{equation}
    \prom{\Hat T}_{tr} =  \frac{1}{2\epsilon}- 
    \frac{m}{2N\epsilon^2}
    \frac{\int   dk_1 dk_2\  g(k_1, k_2) k_2e^{-\epsilon H(k_1, k_2)}}
         {\int   dk_1 dk_2\  g(k_1, k_2) e^{-\epsilon H(k_1, k_2)}},
         \label{equ:promkinek1k2}
  \end{equation}
  where $H(k_1, k_2) = k_1 + mk_2 / 2\epsilon^2$,  
  and $g(k_1, k_2)$ is the number of configurations 
  \mbox{$x=\{x_0, x_1, \dotsc , x_{N-1} \}$} of the classical system with 
  same values of $k_1$ and $k_2$ for fixed $N$.  

  Expressions~\eqref{equ:zetatrk1k2}--\eqref{equ:promkinek1k2} are continuous
  equivalent to those of classical systems~\eqref{equ:classenerprom}. We can use 
  any histogram method
  to compute $g(k_1, k_2)$. Let us use the SHM to illustrate this point.
  First, we fix the simulation temperature
  $T_{sim} = 1/N\epsilon_{sim}$ by choosing $N$ and the initial
  $\epsilon_{sim}$. Then, samples are taken with probabilities proportional
  to $e^{-\epsilon H(x)}$ by using the Metropolis algorithm described above,
  and an histogram of visits $\mathcal{V}(k_1, k_2)$ is cumulated as the
  number of samples with values $k_1$ and $k_2$. Next, the density of
  states $g(k_1, k_2)$ is estimated as
  \begin{equation}\label{equ:quangk1k2shm}
    g(k_1, k_2) \simeq \frac{\mathcal{V}(k_1,
      k_2)}{\mathcal{N}}\exp[\epsilon H(k_1, k_2)].
  \end{equation}
  Once $g(k_1, k_2)$ is computed, averages at different temperatures are
  obtained by changing $\epsilon$ in
  equations~\eqref{equ:prompotek1k2}--\eqref{equ:promkinek1k2}. 
  \begin{figure}
    \centering
    \fbox{
      \includegraphics[scale=0.32]{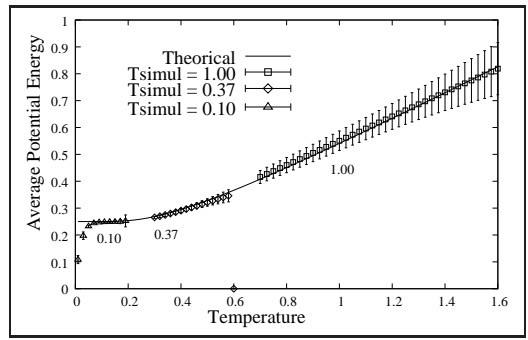}
    }
    \caption{\label{fig:prompotetota}Average potential energy for the
      one--dimensional quantum harmonic oscillator at various simulation
      temperatures.}
  \end{figure}

  \section{\label{sec:shmoacu2} The SHM for the One-Dimensional Quantum
    Harmonic Oscillator}
  The system we chose is a canonical ensemble of
  non-coupled one-dimensional quantum
  harmonic oscillators. For this system, $k_1$
  takes the form $k_1 = \frac{1}{2}m\omega^2 \sum_{l=0}^{n-1}x_l^2$,
  with $m$ the mass  and $\omega$ the
  frequency. The exact expressions for the expectation values
  are~\cite{reichl, reif}
  \begin{equation}\label{equ:prompotekineexac}
    \prom{\Hat H} = 2\prom{\Hat T} = 2\prom{\Hat V} =
    \hbar\omega\left( \frac{1}{2} +
    \frac{1}{e^{\beta\hbar\omega} - 1} \right).
  \end{equation}

  \begin{figure}
    \centering
    \fbox{
      \includegraphics[scale=0.32]{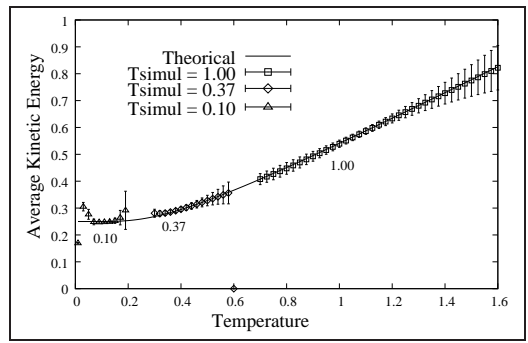}
    }
    \caption{\label{fig:promkinetota}Average kinetic energy for the
      one--dimensional quantum harmonic oscillator at various simulation
      temperatures.}
  \end{figure} 

\begin{figure}
    \centering
    \fbox{
      \includegraphics[scale=0.32]{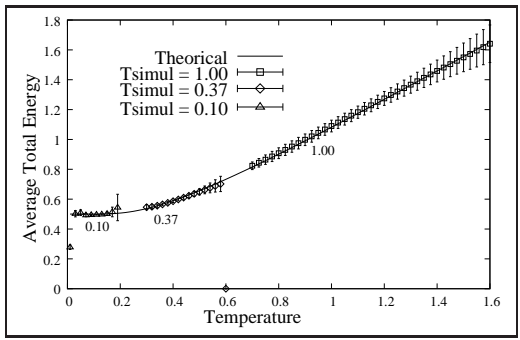}
    }
    \caption{\label{fig:promenertota}Average total energy for the
      one--dimensional quantum harmonic oscillator at various simulation
      temperatures.}
  \end{figure}

  Figures~\ref{fig:prompotetota}, \ref{fig:promkinetota} and
  \ref{fig:promenertota}, shows
  the expectation values of the potential, kinetic and total energy, at
  three different simulation temperatures, $T_{sim}$. The parameters for
  each simulation are shown in table~\ref{tab:parasimu}. In all cases, 
  the total number of samples was 60000, $\kb = 1$ , $30000$ mcss were
  discarded before sampling,   
  and $10$ mcss were performed between successive samples. To make statistic, the
  whole procedure was run ten times with different seeds of the random
  number generator.

  \begin{table}
    \caption{\label{tab:parasimu}Simulation Parameters}
    \begin{ruledtabular}
      \begin{tabular}{lccr}
        $N$&$\epsilon_{sim}$&$T_{sim}$ %
        & $CPUtime$(s)\\
        \hline
        40 & 0.25 & 0.10 & 412.294\\
        30 & 0.09 & 0.37 & 312.607\\
        10 & 0.10 & 1.00 & 117.532\\
      \end{tabular}
    \end{ruledtabular}
  \end{table}

\begin{figure}
    \centering
    \fbox{
      \includegraphics[scale=0.35]{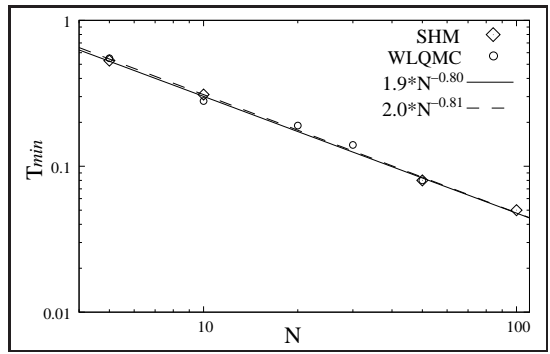}
    }
    \caption{\label{fig:tminshsmwlqc} 
      $T_{min}$ as a function of $N$ for both SHM and WLQMC in
      $\log-\log$ scale.
    }
  \end{figure}

  Errors in the Trotter--Suzuki aproximation are proportional to
  $\epsilon^2$~\cite{lineas}. Since the SHM fixes $N$ and changes
  the temperature by variying $\epsilon$, the errors will be greater
  for lower temperatures. There is a minimum temperature, $T_{min}$,
  where the simulation results includes the theorical ones between the
  error bars. Such a $T_{min}$ is a function of $N$. By looking at
  the total energy, we found for the SHM 
  \begin{equation}\label{equ:tminnshm}
    T_{min} = 1.9(2)N^{-0.80(6)} \quad.
  \end{equation}
  Following the same procedure for the WLQMC we found
  \begin{equation}\label{equ:tminnwlqmc}
    T_{min} = 2.0(6)N^{-0.81(6)} \quad. 
  \end{equation}
  This two results are very similar and are plotted in
  Figure~\ref{fig:tminshsmwlqc}. They actually show that the  minimum
  temperature in  SHM is a consequence of the errors in the 
  Trotter--Suzuki aproximation. 
  This limit is important for the correct performance of the method
  and should
  be taken into account for all applications of histogram methods
  for quantum systems in the WLQMC formulation.
  
  Increasing $N$ or decreasing $\epsilon$ gives exacter results at low
  temperatures, but at price of longer $CPU$ times, just because
  correlation times increases with decreasing $\epsilon$~\cite{lineas}. 
  Nevertheless, the use
  of SHM spares large amounts of $CPU$ times. For example, it takes 118 
  $sec$ on a Pentium 4 at 1.6 GHz to compute $\prom{\Hat H}_T$ over the whole
  temperature range $0.7 \leq T \leq 1.7$, at simulation temperature 
  $T_{sim} = 1.0$, in contrast with the 35 $sec$ of a single point by WLQMC.

  \section{\label{sec:conclu}Conclusions}
  In this work we have employed the extension of histogram methods for
  quantum systems in the WLQMC formulation proposed
  in~\cite{brazilean,josedaniel} to the case of a canonical ensemble of
  one-dimensional quantum harmonic oscillators. By doing so, we have
  explored the ranges of accuracy of the method as a function of the
  number of divisions $N$ in time direction. Actually, since the method
  fixes $N$ and varies $\epsilon$ to change $T$, the corrections in
  the Trotter-Suzuki  expansion fixes a minimum temperature for the
  correct performance of the SHM.
  We obtained that the method
  is accurate for temperatures above $T_{min}$$=$$1.9(2)N^{-0.80(6)}$.
  That is corroborated by direct WLQMC simulations,
  equation~\ref{equ:tminnwlqmc}. 
  This $T_{min}$ can be lowered by increasing $N$, but at cost of
  larger $CPU$ times, because correlation times grow with decreasing
  $\epsilon$. Perhaps SHM needs smaller $CPU$ times than WLQMC.

  The definitions of $k_1$ and $k_2$ and the density of states
  $g(k_1, k_2)$, expressions~\eqref{equ:kunodos} and
  \eqref{equ:quangk1k2shm}, allow to apply any histogram Monte Carlo
  method to quantum systems in the WLQMC formulation, with all the power
  and advantages that histogram methods have deserve for classical ones. 
  However, the restrictions founded here applies to all of them,
  because they relies on the fail of the Trotter-Suzuki expansion for
  lower $\epsilon$ values, and should be always taken into
  account. Further analysis, like the theoretical study of the errors in
  this method, are interesting areas of future work.

  The strategy of extending histogram methods to quantum systems in the
  WLQMC formulation is completely general, and is able to give accurate
  results for many kinds of systems. We expect that our work will
  be useful to the right use of the method for many systems in the future.

  \begin{acknowledgments}
    The authors would like to thank very specially R.~T.~Scalettar for 
    introducing us to Quantum  Monte Carlo and F. Alet for useful
    discussions. Also, J.~D.~Muñoz thanks A. Muramatsu for creative
    discussions on the extension  
    of histogram Monte Carlo methods for quantum systems.
    This research has been supported by the Bogota's Research Division of the
    National University of Colombia (DIB), through the grant 803755.
  \end{acknowledgments}

  \bibliography{biblio}% Produces the bibliography via BibTeX.
\end{document}